\documentclass[12pt]{iopart}
\usepackage{times}
\usepackage{graphicx}
\newcommand{\lab}{\label}

\newcommand{\bc}{\begin{center}}
\newcommand{\ec}{\end{center}}
\newcommand{\be}{\begin{equation}}
\newcommand{\ee}{\end{equation}}
\newcommand{\bea}{\begin{eqnarray}}
\newcommand{\eea}{\end{eqnarray}}
\newcommand{\beq}{\begin{eqalignno}}
\newcommand{\eeq}{\end{eqalignno}}

\def\vec#1{\mbox{\boldmath $#1$\unboldmath}}

\def\vec#1{{\bf #1}}

\def\lab{\label}

\begin{document}

\title{Quantization of scalar fields in curved background, deformed
Hopf algebra and entanglement}
\author{A Iorio, G Lambiase and G Vitiello}
\address{Dipartimento di Fisica ``E.R.Caianiello", Universit\`a di Salerno, 84100
Salerno, Italy,\\ INFN, Gruppo Collegato di Salerno and INFM, Sezione di Salerno}
\begin{abstract}
A suitable deformation of the Hopf algebra of the creation and annihilation operators
for a complex scalar field, initially quantized in Minkowski space--time, induces the
canonical quantization of the same field in a generic gravitational background. The
deformation parameter $q$ turns out to be related to the gravitational field. The
entanglement of the quantum vacuum appears to be robust against interaction with the
environment.
\end{abstract}
% Leave the next line commented out!
% \maketitle

\section{Introduction}

We shortly report on two main results of some recent works \cite{Iorio:2001te, NPB}
on the quantization of a scalar field in curved background: i) a suitable deformation
of the Hopf algebra for a complex scalar operator field, initially quantized in
Minkowski space--time, induces the canonical quantization of the same field in a
generic gravitational background. The deformation parameter $q$ thus turns out to be
related to the gravitational field. ii) The entanglement of the quantum vacuum
appears to be robust against interaction with the environment.

Thermal properties of quantum field theory (QFT) in curved space--time can be derived
in this deformed algebra setting. On the other hand, it is well known the intimate
relationship between space--times with an event horizon and thermal properties
\cite{ISR,Sanchez}. In particular, it has been shown \cite{Sanchez} that global
thermal equilibrium over the whole space--time implies the presence of horizons in
this space--time. We find that the doubling of the degrees of freedom implied by the
coproduct map of the deformed Hopf algebra turns out to be most appropriate for the
description of the modes on both sides of the horizon. The entanglement between inner
and outer particles with respect to the event horizon appears to be rooted in the
background curvature and it is therefore robust against interaction with the
environment.

\section{Quantization and deformed Hopf algebra}
\label{0}

We consider a complex scalar operator field $\phi (x)$, initially quantized in
Minkowski space--time. To study the quantization procedure in curved space--time, we
treat the gravitational field as a classical background. We start with few notions on
the deformation of the Hopf algebra \cite{Celeghini:1991km,Celeghini:1998sy}. We
shall focus on the case of bosons for simplicity.

The coproduct is a homomorphism which duplicates the algebra, ${\Delta}: {\cal A}\to
{\cal A}\otimes {\cal A}$.  The operational meaning of the coproduct is that it
provides the prescription for operating on two modes. Associated to that, there is
the {\it doubling} of the degrees of freedom of the system. Our finding is that in
the presence of a single event horizon such a doubling perfectly describes the modes
on the two sides of the horizon \cite{Iorio:2001te,NPB} (see also
\cite{Martellini:1978sm}).

The bosonic Hopf algebra for a single mode (the case of modes labelled by the
momentum is straightforward), also called $h(1)$, is generated by the set of
operators $\{ a, a^{\dagger},H,N \}$ with commutation relations:
\be [a , a^{\dagger} ] = 2H \, , ~~~[N , a ] = - a \, , ~~~[ N , a^{\dagger} ] =
a^{\dagger} \, , ~~~[ H , \bullet  ] = 0 , \lab{p22} \ee
where  $H$ is a central operator, constant in each representation. The Casimir
operator is given by ${\cal C} = 2NH -a^{\dagger}a$. In ~$h(1)$ the coproduct is
defined by $ \Delta {\cal O} = {\cal O} \otimes {\bf 1} + {\bf 1} \otimes {\cal O}
\equiv {\cal O}^{(+)} + {\cal O}^{(-)}$, where ${\cal O}$ stands for $a$,
$a^{\dagger}$,  $H$ and $N$.  The $q$-deformation of $h(1)$ is the Hopf algebra
$h_{q}(1)$:
\be [a_{q} , a_{q}^\dagger ] = [2H]_{q} \, , ~~~[N ,  a_{q} ] = - a_{q} , ~~~[ N  ,
a_{q}^\dagger ] = a_{q}^\dagger , ~~~[ H , \bullet ] = 0  , \lab{p26} \ee
where $N_{q} \equiv N$, $H_{q} \equiv H$ and $\displaystyle{[x]_{q} = {{q^{x} -
q^{-x}} \over {q - q^{-1}}}}$. The Casimir operator is given by ${\cal C}_{q} =
N[2H]_{q} -a_{q}^{\dagger}a_{q}$.  The coproduct stays the same for $H$ and $N$,
while for $a_{q}$ and $a_{q}^\dagger$ now it changes. In the fundamental
representation, obtained by setting $H = 1/2$, ${\cal C} = 0$, it is written as
\bea \Delta a_{q} &=& a_{q} \otimes {q^{1/2}} + { q^{-1/2}} \otimes a_{q} = a^{(+)}
q^{1/2} + q^{-1/2} a^{(-)} ~,\nonumber \\~~ \Delta a_{q}^{\dagger} &=&
a_{q}^{\dagger} \otimes {q^{1/2}} + {q^{-1/2}} \otimes a_{q}^{\dagger} =
{a^{(+)}}^{\dagger} q^{1/2} +q^{-1/2} {a^{(-)}}^{\dagger} ~, \label{p28} \eea
where self-adjointness requires that $q$ can only be real or of modulus one. In this
representation $h(1)$ and $h_{q}(1)$ coincide. The differences appear in the
coproduct. Note that $[{a^{(\sigma)}} , {a^{(\sigma ')}}^{\dagger} ] = [{a^{(\sigma
)}}, {a^{(\sigma ')}} ] = 0 , ~ \sigma \neq \sigma ' $ with $\sigma \equiv \pm $~.
Now the key point is that, by setting $q = q(\epsilon) \equiv e^{2\epsilon (p)}$,
suitable linear combinations of the deformed copodruct operation (\ref{p28}) (where
the momentum label is introduced) give \cite{Iorio:2001te}:
\bea
d_p^{(\sigma)} (\epsilon) &=& d_p^{(\sigma)} \cosh \epsilon (p) +
{\bar d}_{\tilde p}^{(-\sigma) \dagger} \sinh \epsilon (p)~, \nonumber \\
{\bar d}_{\tilde p}^{(-\sigma) \dagger} (\epsilon) &=& d_p^{(\sigma)} \sinh \epsilon
(p) + {\bar d}_{\tilde p}^{(-\sigma) \dagger} \cosh \epsilon (p)~ ,
\label{b2}
\eea
where $ d_p^{(\sigma)} \equiv \sum_k F (k,p)\, a_k^{(\sigma)}\,, \quad {\bar
d}_p^{(\sigma)} \equiv \sum_k F (k,p)\, {\bar a}_k^{(\sigma)} $ and  $\{F (k,p)\}$ is
a complete orthonormal set of functions \cite{tag}, $p \in {\mathbf{Z}}^{n-1}$, as
for $k = (k_{1}, \vec{k})$, and $p = (\Omega , \vec{p})$~, ${\tilde p} = (\Omega , -
\vec{p})$~. We use $q(p)=q({\tilde p})$~. In general $k \neq p$~. $a_k^{(\sigma)}$
and ${\bar
 a}_k^{(\sigma)}$ are the two (annihilation) operator
modes of the complex scalar field $\phi (x)$ (for each of the sides $\pm$ of the
horizon). Eqs.  (\ref{b2}) are recognized to be the Bogolubov transformations
obtained in the quantization procedure in the gravitational background in the
semiclassical approximation \cite{tag}. We thus see that use of the deformed
coproducts is equivalent to such a quantization procedure.

The generators of  (\ref{b2}) is $ g(\epsilon) =\sum_p \sum_{\sigma} \epsilon(p)
 [d_p^{(\sigma)}\bar{d}_{\tilde p}^{(-\sigma)} -
 d_p^{(\sigma)\,\dagger}\bar{d}_{\tilde
 p}^{(-\sigma)\,\dagger}] $ and $G(\epsilon) \equiv \exp g(\epsilon)$ is a
 unitary operator at finite volume.
The Hilbert--Fock space $\cal  H$ associated to the Minkowski space is built by
repeated action of $(d_p^{(\sigma)\,\dagger}, \bar{d}_{\tilde{p}}^{(-\sigma)\,\dag})$
on the vacuum state $|0_M \rangle$. The generator $G(\epsilon)$ maps vectors of $\cal
H$ to vectors of another Hilbert space ${\cal H}_{\epsilon}$: ${\cal H} \to {\cal
H}_{\epsilon}$. In particular,
\be\label{19} |0(\epsilon) \rangle\,=\,G(\epsilon)\,|0_M \rangle\,{,} \ee
where $|0(\epsilon) \rangle$ is the vacuum state of the Hilbert space ${\cal
H}_{\epsilon}$ annihilated by the new operators ($d_p^{(\sigma)} (\epsilon)$ ,
$\bar{d}_{\tilde p}^{(-\sigma)}(\epsilon)$). We use the short-hand notation for the
Hilbert spaces (${\cal H}$ stands for $\cal H \otimes \cal H$), as well as for the
states (for instance $|0_M \rangle$ stands for $|0_M \rangle \otimes |0_M \rangle$).
The group underlying this construction is $SU(1,1)$. By inverting Eq. (\ref{19}),
$|0_M \rangle$  can be expressed as a $SU(1,1)$ generalized coherent state \cite{PER}
of Cooper-like pairs
\be\label{26}
|0_M \rangle=\frac{1}{Z}\,\exp\left[{\sum_{\sigma} \sum_p
\;\tanh\epsilon (p)
 d_p^{(\sigma)\dagger}(\epsilon) \bar{d}_{\tilde
 p}^{(-\sigma)\dagger}}(\epsilon)\right]\, |0(\epsilon) \rangle\,{,}
 \ee
where $Z= \prod_p\;\cosh^2\epsilon(p)$. Moreover, $\langle 0(\epsilon)|0(\epsilon)
\rangle=1, \forall \epsilon$, and $\langle 0(\epsilon)|0_M \rangle\to 0$ and $\langle
0(\epsilon)|0(\epsilon^{\prime}) \rangle\to 0$ as $V\to\infty, \quad \forall
  \epsilon, \epsilon^{\prime}, \epsilon\ne \epsilon^{\prime}$,
i.e. ${\cal  H}$ and ${\cal H}_\epsilon$ become unitarily inequivalent in the
infinite-volume limit. In this limit $\epsilon$ labels the set $\{H_\epsilon, \forall
\epsilon\}$ of the infinitely many unitarily inequivalent representations of the
canonical commutation relations
\cite{Celeghini:1998sy,Iorio:1993jn,Celeghini:1993jh}.

The physical meaning of having two distinct momenta $k$ and $p$ for states in the
Hilbert spaces ${\cal H}$ and ${\cal H}_\epsilon$, respectively, is the occurrence of
two {\it different} reference frames: the $M$-frame (Minkowski) and the
$M_\epsilon$-frame. To explore the physics in the $M_\epsilon$--frame, one has to
construct a diagonal operator $H_{\epsilon}$ which plays the role of the Hamiltonian
in the $M_\epsilon$-frame. In order to do that one has to use the generator of the
boosts. Thus one finds \cite{Iorio:2001te} \bea
  H_{\epsilon}&=&G(\epsilon){\cal M}_{10}G^{-1}(\epsilon)
   =  \sum_{\sigma}\sum_p \sigma \Omega [ d_p^{(\sigma)
 \dagger}(\epsilon) d_p^{(\sigma)}(\epsilon) +
 \bar{d}_{\tilde{p}}^{(\sigma)}(\epsilon)
 \bar{d}_{\tilde{p}}^{(\sigma)\dagger}(\epsilon)] \nonumber \\
 &=& H^{(+)}(\epsilon) - H^{(-)}(\epsilon) \,{.}
 \label{hamrin}
 \eea
Here ${\cal M}_{10}$ denotes the deformed generator of the boosts. Eq. (\ref{hamrin})
gives the wanted Hamiltonian in the $M_\epsilon$-frame, as also suggested by the
customary results of QFT in curved space-time \cite{tag}.

\section{Entropy and entanglement}

The condensate structure of the vacuum (\ref{26}) suggests to consider the thermal
properties of the system. The entropy operator is $S^{(\sigma)}(\epsilon) = {\cal
S}^{(\sigma)}(\epsilon) + \bar{\cal
 S}^{(\sigma)}(\epsilon)$ with ${\cal S}^{(\sigma)}(\epsilon)$ given by
 ($\sigma \equiv \pm$)
 \be\label{S+}
 {\cal S}^{(\sigma)}(\epsilon)= - \sum_p [d_p^{(\sigma) \dagger}(\epsilon) d_p^{(\sigma)}(\epsilon)
  \ln\sinh^2\epsilon (p)
 - d_p^{(\sigma)}(\epsilon)d_p^{(\sigma)\dagger}(\epsilon) \ln\cosh^2\epsilon
 (p)].
 \ee
$\bar{\cal S}^{(\sigma)}(\epsilon)$ has a similar form (with $d_p \rightarrow {\bar d
}_p$)~. The total entropy operator is  $S_\epsilon =
 S^{(+)}(\epsilon) - S^{(-)}(\epsilon)$ and it is invariant under the Bogoliubov
 transformations.
 Similarly one may introduce the free energy as \cite{11.,FREE}
 \be\label{22}
  {\cal F}^{(+)}(\epsilon)\equiv \langle 0_M|H^{(+)} (\epsilon)
  -\frac{1}{\beta} \, S^{(+)}(\epsilon)|0_M \rangle \,{.}
 \ee
with $\beta \equiv T^{-1}$. Stationarity of ${\cal F}^{(+)}(\epsilon)$ gives
 \be\label{24}
 {\cal N}^{(+)}_{d(\epsilon)} =
 \sinh^2\epsilon (p)= \frac{1}{e^{\beta\Omega}-1}\,{,}
 \ee
and similarly for ${\cal N}^{(+)}_{\bar{d}(\epsilon)}$. Eq. (\ref{24}) shows
that for
vanishing $T$ the deformation parameter $\epsilon$ vanishes too. In that limit
thermal properties as well as the event horizon are lost, and $M_\epsilon$-frame
$\to$ $M$-frame. Moreover, $i)$ $\beta$ is related to the event horizons, and being
$\beta$ constant in time the $M_\epsilon$ space--time is static and stationary; $ii)$
the gravitational field itself vanishes as $\epsilon \to 0$. The
 vanishing of the gravitational field occurs either if the
 $M$-frame is far from the gravitational source where space-time is
 flat, or if there exists a reference frame locally flat, i.e. the
 $M$-frame is a free--falling reference frame. This clearly is a realization
 of the equivalence principle, which manifests itself when
 "$\epsilon$-effects" are shielded.

We now consider the entanglement. The expansion of $|0_M \rangle$ in (\ref{26})
contains terms such as
\be
\sum_p \tanh\epsilon(p) \left( | 1^{(+)}_p , \bar{0} \rangle \otimes |0,
\bar{1}^{(-)}_p \rangle + | 0, \bar{1}^{(+)}_p \rangle \otimes |1^{(-)}_p , \bar{0}
\rangle \right)  + \dots  , \label{expans-vacuum}
\ee
where, we denote by
$|n^{(\sigma)}_p, \bar{m}^{(\sigma)}_p \rangle$ a state of $n$ particles and $m$
``antiparticles" in whichever sector $(\sigma)$. For the generic $n^{\rm th}$ term,
it is  $| n^{(\sigma)}_p , \bar{0} \rangle \equiv |1^{(\sigma)}_{p_1}, \ldots,
1^{(\sigma)}_{p_n}, \bar{0} \rangle$, and similarly for antiparticles. By introducing
a well known notation, $\uparrow$ for a particle, and $\downarrow$ for an
antiparticle, the two-particle state in (\ref{expans-vacuum}) can be written as
\be\label{enta-onestate} | \uparrow^{(+)} \rangle \otimes | \downarrow^{(-)} \rangle
+ | \downarrow^{(+)} \rangle \otimes | \uparrow^{(-)} \rangle \,, \ee
which is an entangled state of particle and antiparticle living in the two sectors
$(\pm)$~. The generic $n^{\rm th}$ term in (\ref{expans-vacuum}) shares exactly the
same property as the two-particle state, but this time the $\uparrow$ describes a
{\it set} of $n$ particles, and $\downarrow$ a {\it set} of $n$ antiparticles. The
mechanism of the entanglement, induced by the q-deformation, takes place at all
orders in the expansion, always by grouping particles and antiparticles into two
sets. Thus the whole vacuum $|0_M \rangle$ is an infinite superposition of entangled
states (a similar structure also arises in the temperature-dependent vacuum of
Thermo-Field Dynamics \cite{11.} (see also \cite{Song})):
\be\label{enta-series}
  |0_M \rangle = \sum_{n=0}^{+\infty} \sqrt{W_n} |{\rm Entangled} \rangle_n
  \,,~~~W_n = \prod_p
  \frac{\sinh^{2n_p}\epsilon(p)}{\cosh^{2(n_p+1)}\epsilon(p)}\,,
\ee
with $ 0< W_n <1 \quad {\rm and} \quad  \sum_{n=0}^{+\infty} W_n = 1$. The
probability of having entanglement of two sets of $n$ particles and $n$ antiparticles
is $W_n$. At finite volume, being $W_n$ a decreasing monotonic function of $n$, the
entanglement is suppressed for large $n$. It appears then that only a finite number
of entangled terms in the expansion (\ref{enta-series}) is relevant. Nonetheless this
is only true at finite volume (the quantum mechanics limit), while the interesting
case occurs in the infinite volume limit, which one has to perform in a QFT setting.

The entanglement is generated by $G(\epsilon)$, where the field modes in one sector
$(\sigma)$ are coupled to the modes in the other sector $(-\sigma)$ via the
deformation parameter $q(\epsilon)$. Since the deformation parameter describes the
background gravitational field (environment), it appears that the origin of the
entanglement {\it is} the environment, in contrast with the usual quantum mechanics
view, which attributes to the environment the loss of the entanglement. In the
present treatment such an origin for the entanglement makes it quite robust. One
further reason for the robustness is that this entanglement is realized in the limit
to the infinite volume {\it once and for all} since then there is no unitary
evolution to disentangle the vacuum: at infinite volume one cannot "unknot the
knots". Such a non-unitarity is only realized when {\it all} the terms in the series
(\ref{enta-series}) are summed up, which indeed happens in the $V\to \infty$ limit
\cite{NPB}.

\end{document}